\documentclass{emulateapj}

\newcommand{\teff}{T$_{eff}$}

\newcommand{\meth}{CH$_4$}
\newcommand{\water}{H$_2$O}

\newcommand{\name}{SDSS~0423$-$0414}

\slugcomment{Accepted to ApJ Letters}

\shorttitle{{\name}AB}
\shortauthors{Burgasser et al.}

\begin{document}

\title{SDSS~J042348.57$-$041403.5AB: A Brown Dwarf Binary Straddling the L/T Transition}

\author{Adam J.\ Burgasser\altaffilmark{1},
I.\ Neill Reid\altaffilmark{2},
S.\ K.\ Leggett\altaffilmark{3},
J.\ Davy Kirkpatrick\altaffilmark{4},
James Liebert\altaffilmark{5}
and
Adam Burrows\altaffilmark{5}}

\altaffiltext{1}{Massachusetts Institute of Technology, Kavli Institute for Astrophysics and Space Research,
77 Massachusetts Avenue, Building 37,
Cambridge, MA 02139-4307, USA; ajb@mit.edu}
\altaffiltext{2}{Space
Telescope Science Institute, 3700 San Martin Drive, Baltimore, MD
21218}
\altaffiltext{3}{Joint Astronomy Centre, 660
North A'ohoku Place, Hilo, HI 96720}
\altaffiltext{4}{Infrared
Processing and Analysis Center, M/S 100-22, California Institute
of Technology, Pasadena, CA 91125}
\altaffiltext{5}{Steward
Observatory, University of Arizona, 933 North Cherry Avenue,
Tucson, AZ 85721}

\begin{abstract}
We present the discovery of SDSS~J042348.57$-$041403.5 as a
closely-separated (0$\farcs$16) brown dwarf binary, resolved by
the $Hubble~Space~Telescope$ Near Infrared Camera and Multi-Object
Spectrometer. Physical association is deduced from the angular
proximity of the components and constraints
on their common proper motion.  {\name}AB appears to be composed of
two brown dwarfs with spectral types L6$\pm$1 and T2$\pm$1.
Hence, this system straddles the transition between L dwarfs and T
dwarfs, a unique evolutionary phase of brown dwarfs
characterized by substantial
shifts in spectral morphology over an apparently narrow
effective temperature range.
Binarity explains a number of unusual
properties of {\name}, including its overluminosity and high effective
temperature compared to other early-type T dwarfs, and possibly
its conflicting spectral classifications (L7.5 in the optical, T0
in the near infrared).
The relatively short estimated orbital period of this system
($\sim$15--20~yr) and the presence
of \ion{Li}{1} absorption in its combined light spectrum make
it an ideal target
for both resolved spectroscopy and dynamical mass measurements.
{\name}AB joins a growing list of late-L/early-T dwarf binaries,
the high percentage of which ($\sim$50\%) may provide
a natural explanation for observed peculiarities across the
L/T transition.
\end{abstract}

\keywords{stars: binaries: visual ---
stars: fundamental parameters ---
stars: individual (\objectname{SDSS J042348.57$-$041403.5}) ---
stars: low mass, brown dwarfs
}

\section{Introduction}

The atmospheres of the lowest-luminosity brown dwarfs,
hosting abundant molecular gas species and
liquid and solid condensates,
have more in common with the atmospheres of giant planets than main sequence
stars. Their unique spectral morphologies have brought about the introduction of
two new spectral classes, L dwarfs and T dwarfs (Kirkpatrick 2005, and references therein).
The former, more luminous objects,
have red near infrared (NIR) colors resulting from warm
photospheric condensate dust clouds;
the latter have blue NIR colors and relatively dust-free photospheres
(e.g., \citet{tsu96b,cha00}).
The transition between these two
classes has turned out to be rather intriguing.
Despite significant evolution in spectral morphology,
effective temperatures ({\teff})
appear to be largely invariant between late-type L dwarfs and mid-type T dwarfs
\citep{gol04}, while
surface fluxes at 1~$\micron$ actually increase
(e.g., \citet{tin03}).
It has been surmised that the L/T transition may be modulated by
a rapid evolution of photospheric condensate dust clouds, as opposed to
the gradual sinking of these clouds as predicted by atmospheric models (e.g., \citet{ack01}).
\citet{me02b} have proposed that these clouds break apart,
creating bright regions analogous to Jupiter's 5 $\micron$ hot spots \citep{wes74}.
\citet{kna04} postulate a phase of ``rapid rainout''.
\citet{tsu03}, however, argue that the
1~$\micron$ brightening arises from variations in
secondary parameters, such as age or metallicity.
With such diversity in its interpretation,
the L/T transition remains an outstanding problem in brown
dwarf astrophysics.

One key source in the midst of this transition is
SDSS~J042348.57$-$041403.5 (hereafter {\name}).
Identified by \citet{geb02} in the Sloan Digital Sky Survey \citep{yor00},
this relatively bright brown dwarf exhibits
weak {\meth} absorption at 1.6 and 2.2 $\micron$ and red NIR colors
indicative of residual photospheric condensate dust.
While these properties are consistent with those of other early-type T dwarfs,
{\name} is noteworthy as its optical spectrum indicates
an earlier L7.5 classification \citep{cru03},
and it is $\sim$1~mag brighter
in the NIR than other similarly-typed dwarfs \citep{vrb04}.
{\name} is also one of the few brown dwarfs to
exhibit both 6708~{\AA} \ion{Li}{1} absorption and 6563~{\AA} H$\alpha$ emission
\citep{me03d}.
It has been surmised that this source could be a young, extremely low
mass brown dwarf; a distended rapid rotator;
or an unresolved binary.
In this Letter, we demonstrate that {\name} is a binary system straddling
the L/T transition.

\section{Observations}

{\name} was observed as part of $Hubble~Space~Telescope$ ($HST$) general observer program 9833,
targeting 22 T dwarfs spanning spectral types T0 to T8
with the Near Infrared Camera and Multi-Object Spectrometer (NICMOS).
Sources were observed with the NIC1 detector
(pixel scale 0$\farcs$0432) and the F110W and F170M filters,
the former sampling the spectral flux peak of
T dwarfs and the latter sampling the 1.6 $\micron$ {\meth}
band.  F110W$-$F170M color measures the depth of
this band, distinguishing T dwarfs from other objects and providing an estimate of their
spectral type.  Full details of our $HST$ program will be presented
in a forthcoming publication (Burgasser et al.\ in prep.).

Observations of {\name} were made on 22 July 2004 (UT). Multiple exposures
totalling 407.7 and 1519.2~s were obtained in the F110W and F170M filters, respectively,
dithering in a spiral pattern with 1$\farcs$3
(30 NICMOS pixels) offsets.
All data were reduced using the STScI calibration pipeline with
CALNICA and the most recent calibration files as of August 2005.
Combined mosaic images were also acquired from the
CALNICB output of the STScI pipeline.

\begin{figure}
\centering
\epsscale{0.4}
\plotone{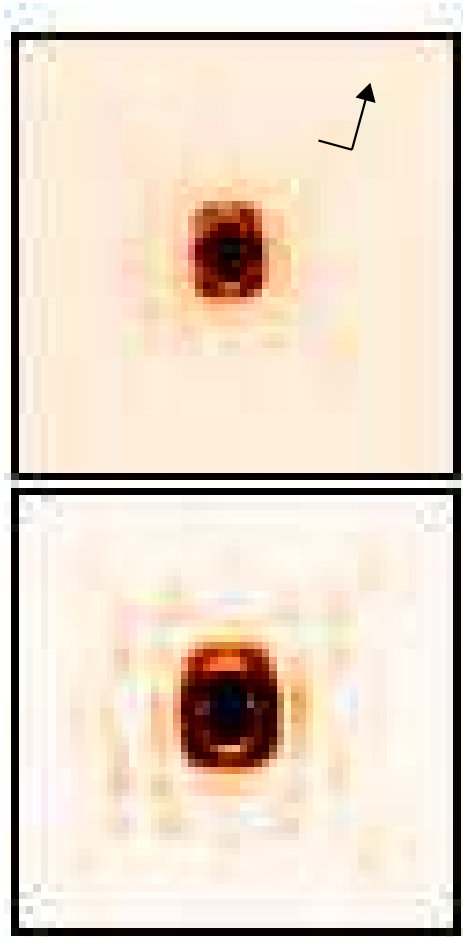}
\caption{$HST$ NICMOS F110W (top) and F170M (bottom) observations of the binary {\name}AB.
Images are 2$\farcs$5 on a side, and
orientation is given by the compass in the top panel indicating north (arrow) and east.
\label{fig1}}
\end{figure}

Figure~\ref{fig1} displays 2$\farcs$5$\times$2$\farcs$5 subsections
of the mosaic images centered on {\name}.  Two overlapping point sources
are resolved, extending along a north-south axis.
Aperture photometry for the combined pair was measured using the
IRAF\footnote{IRAF is distributed by the National Optical
Astronomy Observatories, which are operated by the Association of
Universities for Research in Astronomy, Inc., under cooperative
agreement with the National Science Foundation.} APPHOT routine,
employing a wide (20 pixel radius) circular aperture containing
$\gtrsim$99\% of the total flux of the pair.
Instrumental magnitudes were corrected
to the Arizona Vega system using NICMOS flux calibration parameters;
Vega fluxes of 1785.9 and 946.2 Jy for F110W and F170M, respectively \citep{mob04};
and zeropoints of 0.02 mag.  Uncertainties in the photometry include
photon shot noise and read noise,
0.2\% uncertainty in the NICMOS photometric calibration, 1\% stability
in the zeropoint, and 2\% uncertainty due to the extreme spectral morphology
of the target \citep{mob04}.  Derived magnitudes are listed in Table~1.

To measure the component fluxes, we employed an iterative PSF fitting
routine as described in \citet{me03c}.  Model images were generated from
PSFs of the unresolved sources 2MASS J05591914$-$1404488 and SDSS J125453.90$-$012247.4
\citep[also observed in our $HST$ program]{me00c,leg00}
and compared to individual
calibrated images of {\name}.  A total of 78 fits were made at F110W and 50 fits
at F170M.  Table~1 lists the
means and standard deviations of the angular separation ($\rho$), position angle ($\phi$)
and relative fluxes for the two sources.  The southern source is brighter
by 0.58$\pm$0.11 and 0.85$\pm$0.11 mag at F110W and F170M, respectively; while
the measured separation
($\rho = 0{\farcs}1642{\pm}0{\farcs}0017$)
corresponds to a projected separation of 2.45$\pm$0.07 AU at the distance
of {\name} (15.2$\pm$0.4 pc; Vrba et al.\ 2004).

\section{Analysis}

The colors of the two sources are each consistent
with weak or absent 1.6~$\micron$ {\meth} absorption as expected for a late-type L
or early-type T dwarf.
These colors are also consistent with earlier-type stars, spectral types K--M.
However, the latter are substantially brighter at optical wavelengths
than L or T dwarfs, which have $R-J \gtrsim 6$ \citep{kir99}.
The absence of an optical counterpart at the position of {\name} in
SERC\footnote{Sky Atlas and its Equatorial Extension (SERC)
images were obtained from the Digitized Sky Survey
image server maintained by the Canadian Astronomy Data Centre,
which is operated by the Herzberg Institute of Astrophysics,
National Research Council of Canada.} ER survey images ($R_{limit} \sim 21-22$)
implies $R-J \gtrsim 5-6$ and spectral types
$\gtrsim$M6 for both components.  In addition, no optical counterpart
was detected in deep $R$- and $I$-band images
obtained on 24 January 2003 (UT) using the Palomar
1.5m Telescope facility CCD camera, down to $R \sim 22$ and $I \sim 20$.
We therefore conclude that the two sources are L and/or T dwarfs.

Are they physically associated?  Current estimates of the
surface density of L and T dwarfs for $J \lesssim 16$
(consistent with the brightness of the secondary) are of order 10$^{-3}$ deg$^{-2}$
\citep{me02a,cru03}.
Hence, the probability of two such objects lying within 1$\arcsec$ of
each other is $8{\times}10^{-5}$,
ruling out random alignment at the 99.98\% confidence level.
Common proper motion can be constrained by the fact
that the pair is unresolved in Two Micron All Sky Survey
(2MASS; Cutri et al.\ 2003)
images, for which point sources
can be resolved for separations $\gtrsim$1$\farcs$5 \citep{me05b}.  This restricts
the motion of an unassociated source to 0$\farcs$08--0$\farcs$42 yr$^{-1}$
at a position angle consistent within $\pm$40$\degr$.
These limits, coupled with the late spectrophotometric types and angular
proximity of the two sources, lead us to conclude that they are physically associated.

\begin{figure}
\centering
\epsscale{0.8}
\plotone{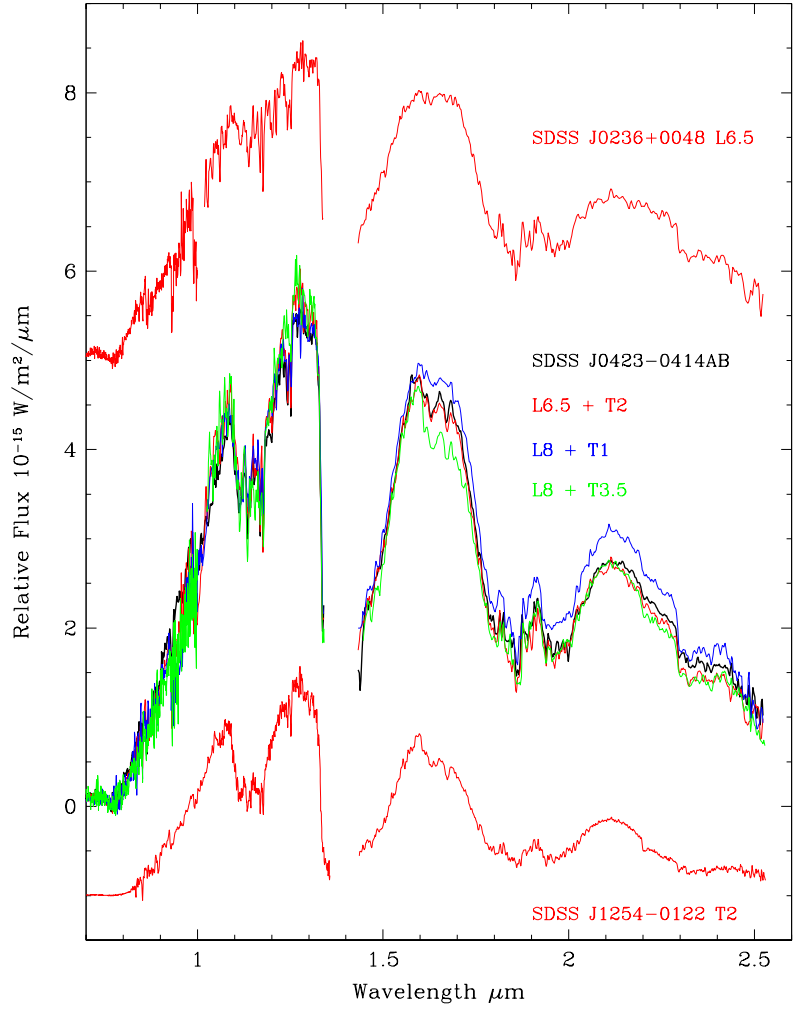}
\caption{Comparison of red optical and near infrared spectral data
for {\name} (center, in black) to various pairings of
late-type L and T dwarf spectra.  The best match of
a combined L6.5 (top) and T2 (bottom) is shown overlain on the spectrum
of {\name} in red.  In contrast, later-type primaries (e.g., L8+T1 in blue)
and secondaries (L8+T3.5 in green) yield poor agreement in band strengths
and/or $H-K$ color.  Earlier-type primaries and later-type secondaries
fail to reproduce the absolute magnitudes of the components.
\label{fig3}}
\end{figure}

The colors of the sources provide only rough constraints on their spectral
types, L5--T3.  Absolute magnitudes also provide weak constraints due
to the inflections in absolute magnitude/spectral type trends
around type T0.
The $M_{F110W}$ and $M_{F170M}$ magnitudes of the brighter (A) component,
14.86$\pm$0.12 and 13.08$\pm$0.12,
are consistent with both mid- and late-L (L5--L7) and
early- and mid-T (T2--T4) types (Burgasser et al., in prep.).
The absolute magnitudes of the fainter (B) component,
15.44$\pm$0.12 and 13.93$\pm$0.12, are similar to
both early- (T0--T1) and mid-type (T4--T5) T dwarfs.
To derive more precise estimates, we compared ``hybrid'' NIR spectra
of late-type L and T dwarf components
to the unresolved spectrum of {\name} \citep[c.f., Cruz et al.\ 2004]{geb02}.
Various combinations of low resolution NIR spectra \citep{leg00,geb02,kna04} for
2MASS~J15074769-1627386 (L5.5\footnote{Spectral types given here are based on the NIR classification
schemes of \citet{geb02} and \citet{me05a}.}),
SDSS~J023617.93+004855.0 (L6.5),
Gliese~584C (L8),
SDSS~J015141.69+124429.6 (T1),
SDSS~J125453.90-012247.4 (T2)
and SDSS~J175032.96+175903.9 (T3.5)
were made after scaling the spectra to match the component
$HST$ fluxes.
Figure~\ref{fig3} demonstrates how an L6.5+T2 combination
provides an excellent match to spectral energy distribution of {\name},
particularly {\meth} and {\water} band strengths.  The hybrid spectrum
has $F110W-F170M = 1.66$, similar to the measured color of 1.69$\pm$0.04.
Good agreements were also found with L5.5+T1, L5.5+T2
and L6.5+T1 combinations.  Using an L8 primary resulted in
{\meth} bands that were too weak and excessive $K$-band flux, while a T3.5 secondary
gave {\meth} bands that were too strong.  Combining these results with
comparisons of the individual colors and absolute magnitudes, we estimate
spectral types
of L6$\pm$1 and T2$\pm$1 for the two components of {\name}, although resolved
spectroscopy is required for more accurate classifications.

The binary nature of {\name}AB explains many of its unusual properties,
most notably its overluminosity and high {\teff} relative to
similarly-classified brown dwarfs.
\citet{gol04} derive $T_{AB}$ = 1450--1825~K from the
combined luminosity,
$\sim$300~K hotter
than typical L8--T2 brown dwarfs.
The component temperatures ($T_A$, $T_B$)
can be deduced from the relative luminosities
and Stefan's Law, assuming similar radii.  However, the luminosity ratio
is difficult to estimate for this system due to the inflection in absolute
brightnesses across the L/T
transition.  If $M_{F110W} \propto L$ then $L_B/L_A$ $\approx$ 0.6; however,
both components could have identical luminosities.
Assuming an average $L_B/L_A$ $\approx$ 0.8, we estimate
$T_A$ = 1250--1575~K and $T_B$ = 1200--1500~K, consistent
with most {\teff}/spectral type trends to date (e.g.,
\citet{gol04,nak04,vrb04}).
We estimate the mass ratio of this pair
to be M$_B$/M$_A \approx (L_B/L_A)^{0.38} = 0.8-1.0$ \citep{bur01}.
The combination of close separation and near-unity mass ratio are common
among substellar field binaries (e.g., \citet{bou03,clo03}).

\section{Discussion}

{\name} joins a growing list of visually resolved substellar binaries,
of which roughly 30 are now known (e.g., Seigler et al.\ 2005).
More intriguing is the addition of another L/T transition binary.
Seven binaries with combined light spectral types spanning L7--T2 have been identified,
including DENIS~J020529.0-115925,
2MASS J17281150+3948593,
2MASS~J21011544+1756586,
DENIS~J225210.73-173013.4,
Gliese~337CD,
2MASS~J05185995-2828372 \citep{bou03,giz03,cru04,me05b,rei05}
and now {\name}.  This is out of the 15 L7--T2 dwarfs so far imaged
at high resolution with $HST$ \citep[Burgasser et al.\ in prep.; Cruz et al.\ in prep.]{rei01,bou03,giz03,rei05},
implying a binary fraction of 47$^{+13}_{-12}$\%.
Even considering biases incurred from the sources having been chosen from
magnitude-limited search samples, this fraction is still over twice that of
magnitude-limited samples of early-type L dwarfs and mid- and late-type T dwarfs
($\sim$20\%; \citet{rei01,giz03,me03c}),
and represents a lower limit
due to the probable existence of tighter, unresolved systems.  An excess
of late-L/early-T binaries could provide a natural explanation for
many of the photometric and spectroscopic peculiarities observed
across the L/T transition.
A more rigorous analysis of binary fraction trends and their implications
will be presented in a forthcoming publication.

The close separation and proximity of the {\name} pair makes it a useful system
for dynamical mass measurements.
Assuming an age of 1--5 Gyr for this system (typical for field sources)
and component {\teff}s as deduced above, evolutionary models
\citep{bur97} predict component masses of 0.040--0.075 M$_{\sun}$;
at least one component has M $\lesssim$ 0.065 M$_{\sun}$
due to the presence of \ion{Li}{1} absorption in the composite optical spectrum.
These estimates imply an orbital period of 15--20~yr (assuming an orbital
separation of 1.29$\rho$; Fischer \& Marcy 1992).  Hence,
$HST$ and/or ground-based adaptive optics observations
should be able to measure significant orbital motion
over the next few years.
Furthermore, \citet{liu05}
have pointed out that binary systems with Li absorption can be age-dated
to higher precision than other field brown dwarfs, particularly
if only one component exhibits the \ion{Li}{1} line.  This system
could therefore be exploited
to empirically test brown dwarf theoretical models with precise mass and age measurements.
Resolved spectroscopy and monitoring observations are needed to
more fully characterize the component brown dwarfs, but such observations
will provide crucial empirical constraints on their physical properties
and the underlying physics of the L/T transition.

\acknowledgments

AJB acknowledges useful discussions with N.\ Siegler and L.\ Close
over binary statistics of late-type dwarfs, and support from NASA through the Spitzer
Fellowship Program.
AB acknowledges support under NASA grant NNG04GL22G.
Based in part on observations made with the
NASA/ESA $Hubble~Space~Telescope$, obtained at the Space Telescope
Science Institute, which is operated by the Association of
Universities for Research in Astronomy, Inc., under NASA contract NAS 5-26555.
These observations are associated with proposal ID GO-9833.
This publication makes use of data from the Two Micron All Sky Survey, which is a
joint project of the University of Massachusetts and the Infrared
Processing and Analysis Center, and funded by the National
Aeronautics and Space Administration and the National Science
Foundation.
2MASS data were obtained from the NASA/IPAC Infrared
Science Archive, which is operated by the Jet Propulsion
Laboratory, California Institute of Technology, under contract
with the National Aeronautics and Space Administration.

\begin{deluxetable}{lll}
\tabletypesize{\footnotesize}
\tablecaption{Properties of SDSS~J042348.57$-$041403.5AB.}
\tablewidth{0pt}
\tablehead{
\colhead{Parameter} &
\colhead{Value} &
\colhead{Ref} \\
}
\startdata
 SpT & L7.5/T0\tablenotemark{a}  & 1,2 \\
     & L6$\pm$1 (A) & 3 \\
     & T2$\pm$1 (B) & 3 \\
 Distance & 15.2$\pm$0.4 pc & 4  \\
 $\mu$ & 0$\farcs$333$\pm$0$\farcs$003 yr$^{-1}$ & 4  \\
 $\theta$ & 284$\fdg$2$\pm$0$\fdg$2 & 4  \\
 $\log{L_{bol}/L_{\sun}}$ & $-$4.14$\pm$0.04\tablenotemark{b} & 5 \\
  & $-$4.39$\pm$0.09 (A) & 3 \\
  & $-$4.50$\pm$0.10 (B) & 3 \\
 {\teff} & 1450--1825~K\tablenotemark{b} & 5 \\
  & 1250--1575~K (A) & 3 \\
  & 1200--1500~K (B) & 3 \\
 F110W & 15.27$\pm$0.03 mag\tablenotemark{b} & 3 \\
 F170M & 13.58$\pm$0.03 mag\tablenotemark{b} & 3 \\
 ${\Delta}$F110W & 0.58$\pm$0.11 mag & 3 \\
 ${\Delta}$F170M & 0.85$\pm$0.11 mag & 3 \\
 $\rho$  & 0$\farcs$1642$\pm$0$\farcs$0017 & 3 \\
  & 2.45$\pm$0.07 AU & 3 \\
  $\phi$ & 20$\fdg$3$\pm$0$\fdg$8 & 3 \\
 M$_{total}$ & 0.08--0.14 & 3,6 \\
 M$_B$/M$_A$ & 0.8--1.0 & 3 \\
 Period & $\sim$15--20 yr & 3 \\
\enddata
\tablenotetext{a}{Optical/NIR spectral type for combined light spectrum.}
\tablenotetext{b}{For combined pair.}
\tablerefs{(1) \citet{cru03}; (2) \citet{geb02}; (3) This paper; (4) \citet{vrb04};
 (5) \citet{gol04}; (6) \citet{bur97}.}
\end{deluxetable}

\end{document}